\documentclass[11pt,letterpaper]{article}

\usepackage[margin=1in]{geometry}
\usepackage[T1]{fontenc}
\usepackage[utf8]{inputenc}
\usepackage{authblk}
\usepackage[numbers,sort&compress]{natbib}

\usepackage{amsmath,amssymb,amsfonts,mathtools}
\usepackage{microtype}
\usepackage{graphicx}
\usepackage{bm}
\usepackage{tikz}
\usepackage{tikz-cd}
\usetikzlibrary{arrows.meta, positioning}
\usepackage{amsthm}
\newtheorem{lemma}{Lemma}
\newtheorem{corollary}{Corollary}

\usepackage{xcolor}
\usepackage[percent]{overpic}
\usepackage[hidelinks]{hyperref}
\usepackage{bbm}

\newcommand{\panelLabel}[1]{%
  {\bfseries\fontsize{9}{9}\selectfont (#1)}%
}

\newcommand{\E}{\mathbb{E}}
\newcommand{\Prb}{\Pr}
\newcommand{\ord}{\operatorname{ord}}
\newcommand{\lcmop}{\operatorname{lcm}}
\newcommand{\Fix}{\operatorname{Fix}}

\begin{document}

\title{Permutation order governs basin entropy in critical one-input Boolean networks}

\author[1]{Ruodan Liu}
\author[1]{Ahana Ghosh}
\author[1]{Sean Campbell}
\author[2]{Venkata Sai Narayana Bavisetty}
\author[3]{Matthew Wheeler}
\author[1]{Claus Kadelka\thanks{Corresponding author: \href{mailto:ckadelka@iastate.edu}{ckadelka@iastate.edu}}}
\affil[1]{Department of Mathematics, Iowa State University, Ames, Iowa 50011, USA}
\affil[2]{Department of Integrative Biology and Physiology, University of California, Los Angeles, California 90024, USA}
\affil[3]{Department of Medicine, University of Florida, Gainesville, Florida 32610, USA}
\renewcommand{\Authfont}{\normalsize}
\renewcommand{\Affilfont}{\small}
\setlength{\affilsep}{0.5em}

\date{November 4, 2026}

\maketitle

\begin{abstract}
Basin entropy measures uncertainty about a network's long-term behavior. 
For every critical one-input synchronous Boolean network with $m$ relevant nodes, we show that the permutation structure induced by its feedback loops determines basin entropy, up to a universal error. 
Although the loops oscillate independently, their relative phases can be offset, producing a deviation from additivity that can grow linearly in $m$.
Yet at the ensemble level, a previously derived additive prediction is preserved through the first subleading term, with the discrepancy appearing at order $\ln m\ln\ln m$.
\end{abstract}

\section*{Introduction}
How does the structure of a dynamical system determine the uncertainty of its long-term behavior? In a finite deterministic system, every trajectory eventually reaches a periodic attractor, and the sizes of the attractors’ basins determine the probabilities of reaching them from uniformly sampled initial states. Boolean networks provide a tractable setting in which to study how network structure shapes this distribution. Introduced by Kauffman as idealized models of genetic regulation~\cite{kauffman1969}, they are now widely used to model gene regulatory systems~\cite{wang2012,davidich2008,kadelka2024}. Under synchronous updating, an $N$-node Boolean network defines a deterministic map $F:\{0,1\}^N\to\{0,1\}^N$. Krawitz and Shmulevich defined basin entropy as Shannon entropy of the attractor reached from a uniformly chosen initial state~\cite{krawitz2007}. That is,
\begin{equation}
H(F)=-\sum_\alpha b_\alpha\ln b_\alpha,
\label{eq:basin-entropy}
\end{equation}
where $b_\alpha$ is the fraction of initial states in the basin of attractor $\alpha$. Relating this global quantity to network structure is challenging because its definition requires the basin-size distribution across an exponentially large state space.

The term basin entropy is also used for continuous dynamical systems, where phase space is partitioned into equal-sized boxes to quantify uncertainty about the final attractor at a specified resolution~\cite{daza2016basin}. 
The basin entropy $H$ used here has no resolution parameter: it is defined on a finite state space under the uniform measure, so uncertainty is determined entirely by basin sizes rather than by the resolution at which they are observed.

In the Kauffman $NK$ model, each of the $N$ nodes has $K$ randomly chosen regulators and a randomly chosen Boolean update function of those regulators. The ordered, critical, and chaotic regimes are distinguished by how small perturbations propagate~\cite{derrida1986}. Critical networks have attracted particular interest because they balance stability and responsiveness~\cite{aldana2007,balleza2008,daniels2018,kadelka2024,kadelka2026canalization}. The critical $K=1$ ensemble is its simplest analytically tractable case and has enabled exact and asymptotic results for attractor numbers and lengths~\cite{flyvbjerg1988,drossel2005,samuelsson2005,fink2023,fink2024,bavisetty2026}. Its long-term dynamics is governed by disjoint feedback loops, each of which propagates binary states by copying or negating its inputs.

Krawitz and Shmulevich found that mean basin entropy grows with network size at criticality, unlike the bounded behavior in the ordered regime and the saturation in the chaotic ensembles they examined~\cite{krawitz2007}. For critical $K=1$ networks, their derivation of the leading growth assumes that the entropies of individual feedback loops are additive. Yet independent loop dynamics do not guarantee additive basin entropy. Under synchronous updating, each loop evolves periodically, and the joint trajectory depends on the loops’ relative phases -- their time offsets with respect to one another. For example, two loops can each have period three, while one loop's state sequence is shifted by one update relative to the other. These phase relationships affect the global attractors and their basin sizes, so independent loop dynamics do not imply additive basin entropy. 

In this Letter, we show that basin entropy is controlled by the order of the permutation, $\sigma$, formed by the feedback loops. For every network with $m$ relevant nodes, we show that
\[H=m\ln 2-\ln\operatorname{ord}(\sigma)+O(1),\] with a universal bound on the error independent of $m$, and derive the sharp bound for each fixed $m$. We also derive the exact entropy composition law and show that the departure from loopwise additivity can be extensive for individual networks. Nevertheless, at the ensemble level, the additive prediction captures the first subleading term, $-\tfrac12(\ln m)^2$, but misses a correction of order $\ln m\ln\ln m$. These results connect a global dynamical entropy to a structural invariant without requiring explicit enumeration of attractors or their basins.

\section*{Expressing basin entropy through state periods}
\begin{figure}[h]
    \centering
    \begin{overpic}[width=0.65\textwidth]{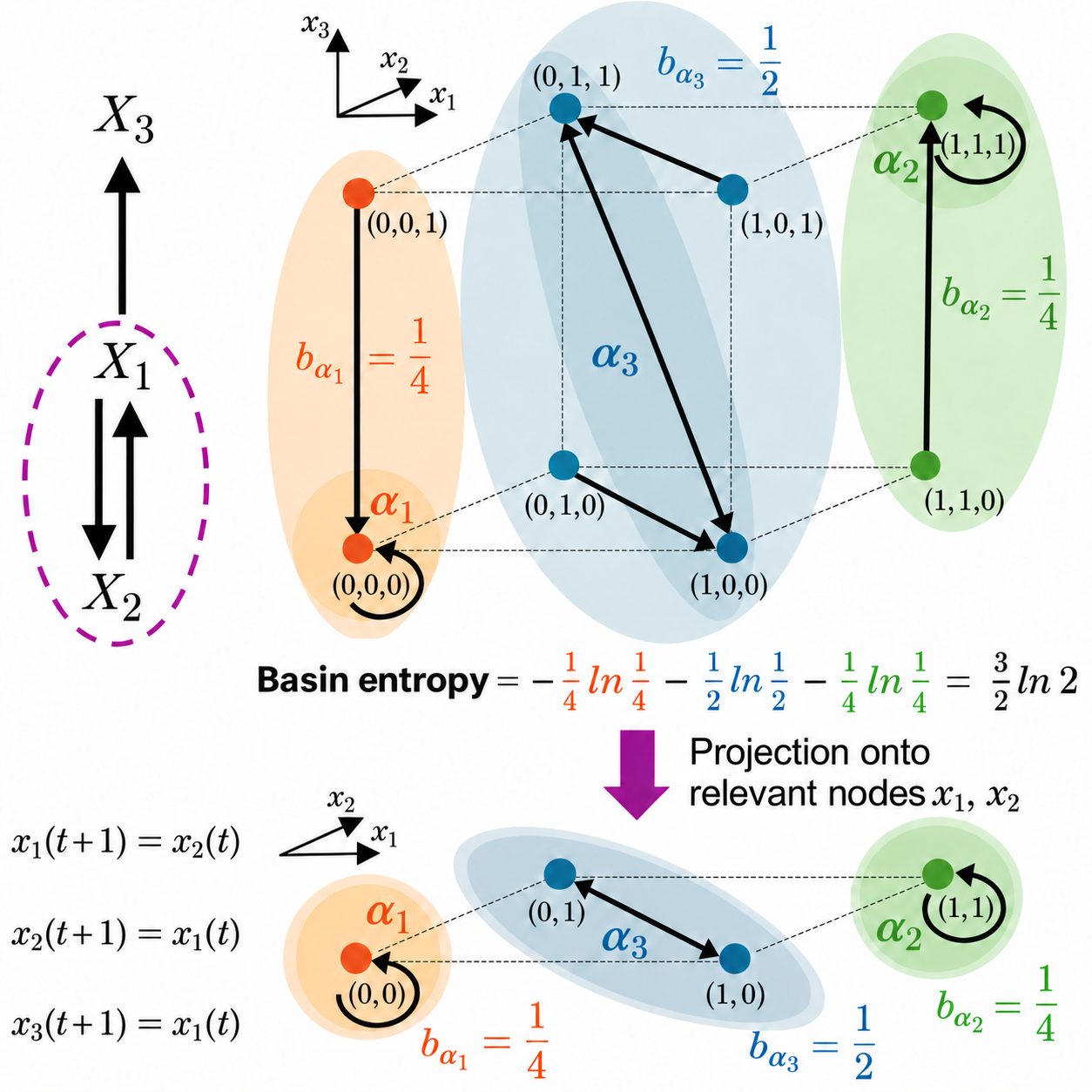}
        \put(-1,96){\panelLabel{a}}
        \put(-1,30){\panelLabel{b}}
        \put(22,96){\panelLabel{c}}
        \put(22,30){\panelLabel{d}}
    \end{overpic}
    \caption{Restriction to the relevant nodes preserves basin entropy.
    (a,b) Wiring diagram and update functions of a three-node critical $K=1$ network with two relevant nodes, $x_1$ and $x_2$, forming a positive loop.
    (c) The synchronous state transition graph has two fixed-point attractors and one period-two attractor, with relative basin sizes $1/4$, $1/4$, and $1/2$.
    (d) Projection onto the relevant nodes preserves attractor periods and relative basin sizes, while eliminating transient states.}
    \label{fig:relevant_node_restriction}
\end{figure}

Consider a critical $K=1$ Boolean network
\[F:\{0,1\}^N\to\{0,1\}^N,\]
in which every node copies or negates its single input.
Each component contains one directed feedback loop, with trees of nodes extending away from it.
A loop is \emph{positive} if it contains an even number of negations and \emph{negative} otherwise.
The nodes on loops are called \emph{relevant nodes} because they fully determine the long-term dynamics; the remaining nodes on the trees affect only transients
(Fig.~\ref{fig:relevant_node_restriction}a,b). 

Let $F_{\rm rel}:\{0,1\}^m\to\{0,1\}^m$ denote the restriction of $F$ to the $m$ relevant nodes.
Each state $\mathbf{x} \in\{0,1\}^m$ has $2^{N-m}$ extensions to the full state space $\{0,1\}^N$, and all these extensions reach the same attractor. Thus, every attractor basin in the full network is larger than its corresponding basin in $F_{\rm rel}$ by the same factor $2^{N-m}$. Relative basin sizes, and hence basin entropy, are unchanged~\cite{flyvbjerg1988}:
\begin{equation}
    H:=H(F)=H(F_{\rm rel}).
\end{equation}
Therefore, basin entropy depends only on the feedback loops
(Fig.~\ref{fig:relevant_node_restriction}c,d).

Because each relevant node copies or negates its input on a feedback loop, $F_{\rm rel}$ is bijective.
Thus, every state $\mathbf{x}\in\{0,1\}^m$ lies on an attractor.
An attractor $\alpha$ of length $|\alpha|$ has relative basin size $b_\alpha=|\alpha|/2^m$.
Choose $X$ uniformly from $\{0,1\}^m$, and let $T(X)$ denote the period of the attractor containing $X$.
Then Eq.~\eqref{eq:basin-entropy} becomes
\begin{equation}
H=m\ln 2-\mathbb E_X[\ln T(X)].
\label{eq:entropy-period}
\end{equation}

\section*{Nonadditivity across loops}

The period representation of basin entropy shows why contributions from separate loops need not add. 
Suppose the network has $k$ disjoint feedback loops, with loop $i$ containing $m_i$ relevant nodes, so that $\sum_{i=1}^k m_i=m$. A uniformly sampled state of the relevant nodes can then be written as $X=(X_1,\ldots,X_k)$, where $X_i\in\{0,1\}^{m_i}$ is the state on loop $i$. Let $T(X_i)$ denote the period of the attractor containing $X_i$. Under synchronous updating, the period of the joint state is
\begin{equation}
T(X)=\operatorname{lcm}\bigl(T(X_1),\ldots,T(X_k)\bigr).
\label{eq:global-state-period}
\end{equation}
Let $H_i$ denote the basin entropy of loop $i$ in isolation.
Applying Eq.~\eqref{eq:entropy-period} to the full network and to its individual loops gives the exact composition law
\begin{equation}
\begin{aligned}
H-\sum_{i=1}^k H_i
&=\sum_{i=1}^k\mathbb E_X[\ln T(X_i)]
  -\mathbb E_X[\ln T(X)]\\
&=\mathbb E_X\!\left[
\ln\frac{\prod_{i=1}^kT(X_i)}
{\operatorname{lcm}(T(X_1),\ldots,T(X_k))}
\right].
\end{aligned}
\label{eq:composition-law}
\end{equation}
For a fixed choice of one attractor on each loop, let $t_i$ be the period of the chosen attractor on loop $i$. Their Cartesian product contains $\prod_i t_i$ joint states. Under synchronous updating, each resulting global attractor has period $\operatorname{lcm}(t_1,\ldots,t_k)$, so the number of global attractors formed from this choice is
\[
\frac{\prod_i t_i}{\operatorname{lcm}(t_1,\ldots,t_k)}.
\]
Thus, the ratio in Eq.~\eqref{eq:composition-law} counts the global attractors associated with the selected loop attractors. It is large when the loop periods share prime factors: the product counts all combinations of loop states, while their shared factors shorten the period of each global attractor. The logarithm of this count is the additional uncertainty about the global attractor beyond that captured by the separate loop entropies.

The discrepancy in Eq.~\eqref{eq:composition-law} can be severe.
Consider $k$ disconnected negative loops of length two.
Each loop has a single period-four attractor and hence $H_i=0$.
The combined network has $4^k$ states, partitioned into $4^{k-1}$ period-four attractors of equal basin size, giving $H=(k-1)\ln4$.
More generally, for fixed $m$,
\begin{equation}
0\le H-\sum_i H_i\le(m-1)\ln2,
\label{eq:nonadd-worst}
\end{equation}
with equality attained by $m$ negative self-loops (see End Matter).

\begin{figure}[!htb]
    \centering
    \begin{overpic}[width=0.65\textwidth]{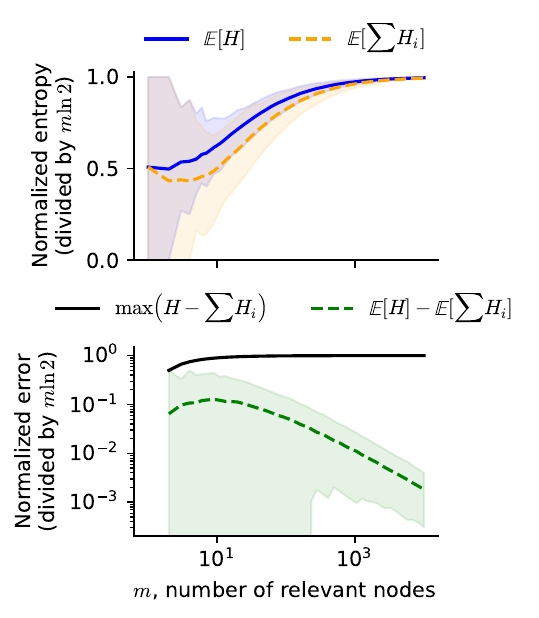}
        \put(-1,96){\panelLabel{a}}
        \put(-1,48){\panelLabel{b}}
    \end{overpic}
    \caption{Mean and worst-case effects of nonadditivity.
    (a) Ensemble means of $H/(m\ln2)$ (blue solid) and $\sum_i H_i/(m\ln2)$ (orange dashed).
    (b) Ensemble mean of $(H-\sum_i H_i)/(m\ln2)$ (green dashed) and its exact worst-case value, $1-1/m$ (black solid), attained by negative self-loops.
    Shading shows the empirical 2.5\%--97.5\% percentiles across sampled collections of feedback loops with $m$ total nodes and independent, equiprobable loop signs. Each panel uses 3,000 such samples for each of 50 logarithmically spaced values of $m$ between 1 and 10,000. The $m=1$ value is omitted from (b) because the discrepancy is zero.}
    \label{fig:additivity_error}
\end{figure}

Figure~\ref{fig:additivity_error} compares the additive estimate with the basin entropy across randomly sampled collections of feedback loops containing $m$ total relevant nodes, with loop signs chosen independently and equiprobably. The mean discrepancy, normalized by $m\ln 2$, becomes small as $m$ grows, although the discrepancy for individual networks can remain proportional to $m$. To determine the ensemble mean beyond this leading scale, we must analyze the global state period $T(X)$, rather than only the individual loop periods $T(X_i)$. In particular, we next bound the ratio between the maximal possible period $\ell_{\max}$ and the period $T(X)$ of a uniformly sampled state.

\section*{From network structure to basin entropy}
Recall that $m_1,\ldots,m_k$ denote the lengths of the feedback loops. The loops are the disjoint cycles of a permutation $\sigma$ of the $m$ relevant nodes~\cite{bavisetty2026}. Its order -- the smallest positive integer $q$ for which $\sigma^q$ is the identity -- is
\begin{equation}
\operatorname{ord}(\sigma)=\operatorname{lcm}(m_1,\ldots,m_k).
\label{eq:permutation-order}
\end{equation}
Let $\varepsilon_i=0$ for a positive loop and $\varepsilon_i=1$ for a negative loop. The maximal period on loop $i$ is $\ell_i=m_i2^{\varepsilon_i}$,
so the maximal period of the relevant-node dynamics is~\cite{flyvbjerg1988}
\begin{equation}
\ell_{\max}=\operatorname{lcm}(\ell_1,\ldots,\ell_k)
\in\{\ord(\sigma),2\ord(\sigma)\}.
\label{eq:max-period}
\end{equation} 
For a uniformly sampled relevant-node state $X$, the ratio $\ell_{\max}/T(X)$ compares the longest possible attractor period with the period of the attractor containing $X$. This ratio is determined by which prime-power factors of $\ell_{\max}$ are absent from $T(X)$. Our main result bounds its expected logarithm, uniformly over all critical $K=1$ networks:
\begin{equation}
0\leq
\mathbb E_X\!\left[\ln\frac{\ell_{\max}}{T(X)}\right]
\leq \mathfrak c_* =
\sum_{p\ \mathrm{prime}}\frac{\ln p}{2^{p-1}}
\approx 0.755245.
\label{eq:main-bound}
\end{equation}
Thus, although some states can have periods far below \(\ell_{\max}\), the prime-power factors missing from \(T(X)\) have a universally bounded expected logarithmic contribution, independent of the number, lengths, or signs of the loops (proof in the End Matter).

\begin{figure}[!htb]
    \centering
    \begin{overpic}[width=0.65\textwidth]{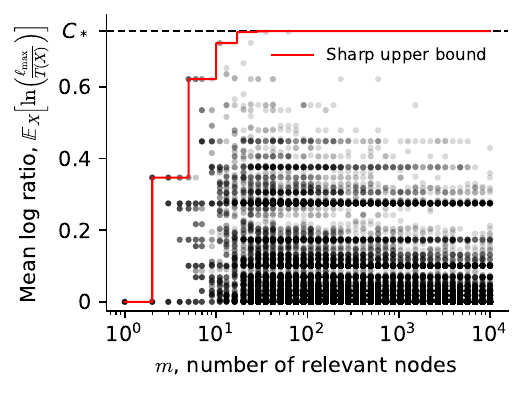}
        \put(-1,77){\panelLabel{a}}
    \end{overpic}

    \vspace{2mm}

    \begin{overpic}[width=0.65\textwidth]{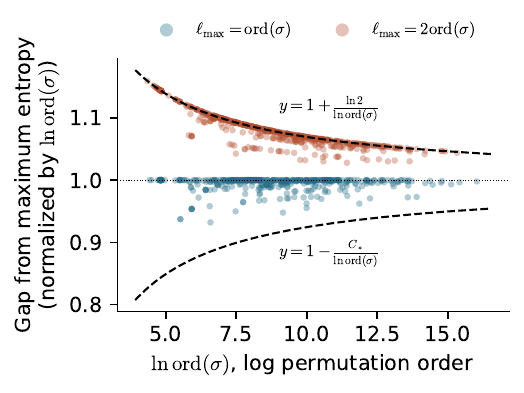}
        \put(-1,71){\panelLabel{b}}
    \end{overpic}

    \caption{Uniform control of basin entropy.
    (a) Mean log ratio $\mathbb E_X\!\left[\ln\frac{\ell_{\max}}{T(X)}\right]$
    for sampled permutations with independent, equiprobable loop signs. Each point represents a distinct observed pair; opacity indicates multiplicity.
    The red staircase gives the sharp bound at fixed $m$ (Corollary~\ref{cor:sharp-fixed-m}), approaching the universal bound $\mathfrak c_*$ (horizontal dashed line).
    (b) Normalized entropy deficit
    $(m\ln2-H)/\ln\ord(\sigma)$ at $m=128$.
    Blue and orange points correspond to
    $\ell_{\max}=\ord(\sigma)$ and
    $\ell_{\max}=2\ord(\sigma)$, respectively.
    The dotted horizontal line marks unity. We sample 3,000 permutations per $m$ in (a) and 1,000 in total in (b).}
    \label{fig:entropy_bounds}
\end{figure}

The bound is sharp. For each fixed $m$, the exact upper bound is attained by positive loops whose lengths are consecutive primes starting from $2$, with any remaining nodes forming positive self-loops. As $m$ increases, these sharp bounds approach $\mathfrak c_*$ (see Fig.~\ref{fig:entropy_bounds}a and Corollary~\ref{cor:sharp-fixed-m} in the End Matter).
Combining Eqs.~\eqref{eq:entropy-period} and
\eqref{eq:main-bound} gives the explicit bound
\begin{equation}
m\ln2-\ln\ell_{\max}
\leq H
\leq m\ln2-\ln\ell_{\max}+\mathfrak c_*.
\label{eq:entropy-signed-order}
\end{equation}
Since $\ell_{\max}\in\{\operatorname{ord}(\sigma),2\operatorname{ord}(\sigma)\}$, this yields
\begin{equation}
H=m\ln2-\ln\operatorname{ord}(\sigma)+O(1),
\label{eq:entropy-order}
\end{equation}
where the remainder is bounded uniformly over all networks, independently of $m$
(Fig.~\ref{fig:entropy_bounds}b).
Thus, the feedback-loop structure determines basin entropy to within a universal absolute error.

Taking the expectation over all possible feedback-loop structures, i.e., uniformly over all permutations in $S_m$ with independent and equiprobable loop signs~\cite{flyvbjerg1988}, yields
\begin{equation}
m\ln2-\mu_m-\ln2
\leq \mathbb{E}_{\sigma,\varepsilon}[H]
\leq m\ln2-\mu_m+\mathfrak c_*.
\label{eq:fixed-core-bound}
\end{equation}
The known asymptotic expansion~\cite{zacharovas2004}
\[
\mathbb{E}_\sigma[\ln\ord(\sigma)]
=\frac12(\ln m)^2-\ln m\ln\ln m+O(\ln m)
\]
therefore yields
\begin{equation}
\mathbb{E}_{\sigma,\varepsilon}[H]
=m\ln2-\frac12(\ln m)^2
+\ln m\ln\ln m+O(\ln m).
\label{eq:fixed-core-asymptotic}
\end{equation}

\section*{Ensemble entropy}
These results have direct implications for the critical $K=1$ ensemble on $N$ nodes, where only a subset of nodes is relevant. For a random network in this ensemble, let $M$ be the number of relevant nodes. Its distribution is known~\cite{flyvbjerg1988}. For $m=1,\ldots,N$, 
\begin{equation}
\Prb(M=m \mid N)
=\frac{m}{N}
\prod_{j=1}^{m-1}\left(1-\frac{j}{N}\right).
\label{eq:core-size-distribution}
\end{equation}
The moment estimates for $M$, derived in the End Matter, then give
\begin{equation}
\begin{aligned}
\E_N[H]
={}&\sqrt{\frac{\pi N}{2}}\,\ln2
-\frac18(\ln N)^2\\
&+\frac12\ln N\ln\ln N+O(\ln N).
\end{aligned}
\label{eq:critical-asymptotic}
\end{equation}

\section*{Assessing loopwise additivity}
To assess the loopwise-additive prediction, we compute its ensemble mean (derivation in End Matter):
\begin{equation}
\mathbb E_{\sigma,\varepsilon}\!\left[\sum_{i=1}^k H_i\right]
=m\ln2-\frac12(\ln m)^2+O(\ln m).
\label{eq:additive-mean}
\end{equation}
Comparing this with Eq.~\eqref{eq:fixed-core-asymptotic} gives
\begin{equation}
\mathbb E_{\sigma,\varepsilon}\!\left[H-\sum_{i=1}^k H_i\right]
=\ln m\ln\ln m+O(\ln m).
\label{eq:mean-nonadditivity}
\end{equation}
Thus, loopwise additivity captures the leading and first subleading terms of the mean basin entropy, but introduces an error of order $\ln m\ln\ln m$.

\section*{Discussion}
Krawitz and Shmulevich derived the leading scaling of ensemble-mean basin entropy using a loopwise-additive calculation~\cite{krawitz2007}.
We show that the discrepancy from additivity can be extensive for individual networks, although ensemble averaging suppresses the discrepancy: for the fixed-$m$ ensemble, the additive approximation remains correct through the first subleading term, with the mean discrepancy first appearing at order $\ln m\ln\ln m$.
More strongly, every critical $K=1$ network satisfies
$H=m\ln2-\ln\ord(\sigma)+O(1)$,
with an error bounded uniformly over network size and loop configuration.
Thus, basin entropy can be estimated from feedback-loop structure without enumerating attractors or their basin sizes.

Extending this description beyond critical $K=1$ networks remains open.
In the critical ensembles studied in Ref.~\cite{mihaljev2006}, all but a finite number of relevant components are asymptotically simple loops.
This makes loops a natural starting point, but does not establish that they determine basin entropy.
Components containing nodes with multiple relevant inputs can have transient states, so their basin sizes are not determined by attractor periods alone.
A broader theory must account for these component-level basin structures and their synchronous composition, and determine whether a similarly simple structural quantity controls the resulting basin entropy.

\section*{Acknowledgements}
This research was supported in part by grants from the NSF (DMS-2235451) and Simons Foundation (MPS-NITMB-00005320) to the NSF-Simons National Institute for Theory and Mathematics in Biology (NITMB). R.L., S.C., and C.K. were further supported by the NSF (DMS-2424632 and DMS-2451973). M.W. was further supported by the NSF (DMS-2424635) and the NIH (R01-AI135128 and R01-HL169974-01). 

\section*{Data availability}
The Python source code used to reproduce Figs.~\ref{fig:additivity_error} and \ref{fig:entropy_bounds}
is openly available at
\url{https://github.com/RuodanL/basin-entropy-critical-k1}.

\section*{Appendix}
\setcounter{equation}{0}
\renewcommand{\theequation}{A\arabic{equation}}
\renewcommand{\theHequation}{A.\arabic{equation}}

We provide the proofs and estimates not developed in the letter. We retain its notation. Specifically, $F_{\rm rel}$ denotes the network restricted to the $m$ relevant nodes, $m_1,\ldots,m_k$ are the loop lengths, and $\ell_i=m_i2^{\varepsilon_i}$ is the maximal period on loop $i$. Subscript $X$ on expectations refer to a uniformly sampled state of relevant nodes.

\subsection*{Worst-case departure from additivity}
The upper bound in Fig.~\ref{fig:additivity_error}b follows from Eq.~\eqref{eq:composition-law}. For every state $X$ and every loop $j$, $T(X_j)$ divides the least common multiple in the denominator. Therefore,

\begin{equation}
\frac{\prod_{i=1}^kT(X_i)}
{\lcmop(T(X_1),\ldots,T(X_k))}
\leq\prod_{i\ne j}T(X_i)
\leq 2^{m-m_j}.
\end{equation}

\noindent Since $m_j\geq1$, the ratio is at most
$2^{m-1}$. Taking logarithms and expectations yields Eq.~\eqref{eq:nonadd-worst}.
The upper bound is attained when all $m$ loops are negative self-loops. In that case the ratio equals $2^{m-1}$ for every state.

\subsection*{Proof of the universal period-deficit bound}
For each prime $p$, let $v_p(n)$ denote its exponent in the prime
factorization of $n$. We first bound the probability that a loop state
or a network state has a period that omits a prime-power factor of the maximal period.

\begin{lemma}\label{lem:single-loop-loss}
For any loop $i$ with maximal period $\ell_i$ and any prime $p\mid\ell_i$, let $a=v_p(\ell_i)$. Choosing $X_i$ uniformly from $\{0,1\}^{m_i}$, then
$\Pr\left(p^a\nmid T(X_i)\right)
\leq 2^{-(p-1)p^{a-1}}$.
If loop $i$ is negative and $p=2$, this probability is zero.
\end{lemma}

\begin{proof}
Let $F_i$ denote the restriction of $F_{\rm rel}$ to loop $i$. By
Ref.~\cite{samuelsson2005}, for every positive integer $t$,

\begin{equation}
|\Fix(F_i^t)|=
\begin{cases}
2^{\gcd(t,m_i)},&
\text{if }\varepsilon_i=0
\text{ or }\dfrac{t}{\gcd(t,m_i)}\text{ is even},\\[4pt]
0,&\text{otherwise}.
\end{cases}
\label{eq:fixedpointsindividualloops}
\end{equation}

Here $\Fix(F_i^t)$ is the set of states $\mathbf x\in\{0,1\}^{m_i}$
satisfying $F_i^t(\mathbf x)=\mathbf x$.
Since $p\mid\ell_i$, set $t=\ell_i/p$. Because every state's period divides $\ell_i$, 
$p^a\nmid T(X_i)$ if and only if $T(X_i)$ divides $t$, therefore
$\Pr\bigl(p^a\nmid T(X_i)\bigr) ={|\Fix(F_i^t)|}/{2^{m_i}}$.
If the loop is positive, or if $p$ is odd, Eq.~\eqref{eq:fixedpointsindividualloops} yields
$|\Fix(F_i^t)|=2^{m_i/p}$. In these cases $p^a$ divides $m_i$, so

\begin{equation}
\Prb\bigl(p^a\nmid T(X_i)\bigr)
=2^{-m_i(1-1/p)}
\leq2^{-(p-1)p^{a-1}}.
\end{equation}

The only remaining case is a negative loop with $p=2$. Then $t=m_i$, so
$t/\gcd(t,m_i)=1$ and the fixed-point count is zero.
\end{proof}

\begin{lemma}\label{lem:one-prime}
Let $\ell_{\max}$ be the maximal period of the relevant-node dynamics, and let $X=(X_1,\ldots,X_k)$ be uniformly sampled from $\{0,1\}^m$. For any prime $p\mid\ell_{\max}$, let
$a_p=v_p(\ell_{\max})$. Then,

\begin{equation*}
\sum_{a=1}^{a_p}\Pr\bigl(p^a\nmid T(X)\bigr)
\leq 2^{-(p-1)}.
\end{equation*}
\end{lemma}

\begin{proof}
Choose a loop $j$ with $v_p(\ell_j)=a_p$. Since $X$ is uniform on $\{0,1\}^m$, $X_j$ is uniform on $\{0,1\}^{m_j}$. For
$1\leq a\leq a_p$, Eq.~\eqref{eq:global-state-period} implies

\[
\Pr\bigl(p^a\nmid T(X)\bigr)
\leq \Pr\bigl(p^a\nmid T(X_j)\bigr)
\leq \Pr\bigl(p^{a_p}\nmid T(X_j)\bigr).
\]

\noindent Lemma~\ref{lem:single-loop-loss} therefore gives

\[
\sum_{a=1}^{a_p}\Pr\bigl(p^a\nmid T(X)\bigr)
\leq a_p\,2^{-(p-1)p^{a_p-1}}
\leq 2^{-(p-1)},
\]

\noindent where the last inequality follows because
$n\,2^{-(p-1)p^{n-1}}$ is nonincreasing for positive integers $n$.
\end{proof}

\begin{proof}[Proof of Eq.~\eqref{eq:main-bound}]
Since $T(X)\mid\ell_{\max}$, prime factorization gives

\begin{equation}
\ln\frac{\ell_{\max}}{T(X)}
=\sum_{p\mid\ell_{\max}}
\bigl(a_p-v_p(T(X))\bigr)\ln p.
\end{equation}

\noindent Moreover,
$a_p-v_p(T(X))
=\sum_{a=1}^{a_p}\mathbbm{1}_{\{p^a\nmid T(X)\}}$. Taking expectations yields

\begin{equation}
\mathbb E_X\!\left[\ln\frac{\ell_{\max}}{T(X)}\right]
=\sum_{p\mid\ell_{\max}}(\ln p)
\sum_{a=1}^{a_p}\Pr\bigl(p^a\nmid T(X)\bigr).
\label{eq:end-prime-decomposition}
\end{equation}

\noindent With Lemma~\ref{lem:one-prime},

\[
\mathbb E_X\!\left[\ln\frac{\ell_{\max}}{T(X)}\right]
\leq\sum_{p\mid\ell_{\max}}\frac{\ln p}{2^{p-1}}
\leq\sum_{p\ {\rm prime}}\frac{\ln p}{2^{p-1}}
=\mathfrak c_*.
\]

\noindent The series converges by comparison with
$\sum_{n\geq2}n/2^{n-1}$.
\end{proof}

\subsection*{Sharpness at fixed number of relevant nodes}
The universal bound in Eq.~\eqref{eq:main-bound} can be sharpened for
each fixed $m$. This gives the staircase in Fig.~\ref{fig:entropy_bounds}a
and shows that $\mathfrak c_*$ is the best possible uniform bound.

\begin{corollary}
\label{cor:sharp-fixed-m}
Let $p_1<p_2<\cdots$ be the primes, and set $s_0=0$ and
$s_q=\sum_{j=1}^q p_j$. If $s_q\leq m<s_{q+1}$, then

\begin{equation}
0\leq\mathbb E_X\!\left[\ln\frac{\ell_{\max}}{T(X)}\right]
\leq\sum_{j=1}^q\frac{\ln p_j}{2^{p_j-1}}.
\label{eq:end-fixed-m-sharp}
\end{equation}

\noindent The upper bound is attained by one positive loop of each length
$p_1,\ldots,p_q$, with the remaining $m-s_q$ nodes forming positive
self-loops.
\end{corollary}

\begin{proof}
Write $L=\ord(\sigma)$. Lemma~\ref{lem:one-prime} and
Eq.~\eqref{eq:end-prime-decomposition} imply

\begin{equation}
\mathbb E_X\!\left[\ln\frac{\ell_{\max}}{T(X)}\right]
\leq\sum_{p\mid L}\frac{\ln p}{2^{p-1}}.
\label{eq:end-primes-of-L}
\end{equation}

\noindent Indeed, since $\ell_{\max}\in\{L,2L\}$, the only possible prime factor
of $\ell_{\max}$ that does not divide $L$ is $2$, and this occurs when
$L$ is odd. Then at least one loop is negative and has odd length.
By Lemma~\ref{lem:single-loop-loss}, every state on that loop has an
even period. Hence every global period $T(X)$ is even, so this extra
factor of $2$ contributes zero to
Eq.~\eqref{eq:end-prime-decomposition}.

The sum of the distinct prime divisors of an integer $n$ is at most
$n$. Therefore,

\[
\sum_{p\mid L}p
\leq\sum_{i=1}^k\sum_{p\mid m_i}p
\leq\sum_{i=1}^k m_i=m.
\]

\noindent Let $q_1<\cdots<q_r$ be the distinct prime divisors of $L$, and let $p_1<p_2<\cdots$ denote the primes in increasing order, as in the corollary. Then $q_j\geq p_j$, so $s_r\leq m$ and therefore $r\leq q$. The function $\ln p/2^{p-1}$ decreases as $p$ increases over the primes: the exponential denominator grows faster than the logarithmic numerator.
Thus,

\[
\sum_{p\mid L}\frac{\ln p}{2^{p-1}}
\leq\sum_{j=1}^r\frac{\ln p_j}{2^{p_j-1}}
\leq\sum_{j=1}^q\frac{\ln p_j}{2^{p_j-1}}.
\]

To attain equality, take one positive loop of each length $p_1,\ldots,p_q$. On a positive loop of prime length $p_j$, the two constant states have period $1$, while all other states have period $p_j$. No other loop supplies the factor $p_j$, so $\Pr\bigl(p_j\nmid T(X)\bigr)=2^{1-p_j}$. Equation~\eqref{eq:end-prime-decomposition} then gives equality in Eq.~\eqref{eq:end-fixed-m-sharp}. Adding positive self-loops changes neither $\ell_{\max}$ nor the distribution of $T(X)$. When $m=1$, $q=0$ and the sum is empty. Finally, as $m\to\infty$, the attained bounds converge to $\mathfrak c_*$, proving that the universal constant is optimal.
\end{proof}

\subsection*{Averaging over the number of relevant nodes}
To derive Eq.~\eqref{eq:critical-asymptotic}, we average Eq.~\eqref{eq:fixed-core-asymptotic} over the random number $M$. This requires asymptotic estimates for $\mathbb E[M]$ and the logarithmic moments of $M$. The leading term of $\mathbb E[M]$ is known from Ref.~\cite{flyvbjerg1988}, while a more precise asymptotic expansion in Ref.~\cite{flajolet1995} gives $\mathbb E[M]=\sqrt{\pi N/2}-1/3+O(N^{-1/2})$.

\noindent For the logarithmic moments, set $Y_N=M/\sqrt N$. We first show that $\ln Y_N$ has a uniformly bounded second moment. Using the tail-integral identity $\E[|\ln Y_N|^2]
=2\int_0^\infty t\,\Pr(|\ln Y_N|>t)\,dt$, together with $\Pr(|\ln Y_N|>t)=\Pr(\ln Y_N<-t)+\Pr(\ln Y_N>t)$, it suffices to control the lower and upper tails of $\ln Y_N$. Equation~\eqref{eq:core-size-distribution} gives

\begin{equation}
\Pr(M\geq m)=\prod_{j=1}^{m-1}\left(1-\frac{j}{N}\right), \qquad 1\leq m\leq N.
\label{eq:end-core-tail}
\end{equation}

\noindent Together with $\Pr(M=m)\leq m/N$, this yields

\begin{equation}
\begin{split}
\Pr(M\geq m)&\leq
\exp\!\left[-{m(m-1)}/{2N}\right], \\ 
\Pr(M\leq m)&\leq{m(m+1)}/{2N}.
\label{eq:end-M-tails}
\end{split}
\end{equation}

\noindent For the lower tail, let
$r=\lfloor\sqrt N e^{-t}\rfloor$. If $r\geq1$, the bound on $\Pr(M\leq m)$ in Eq.~\eqref{eq:end-M-tails} gives

\[
\Pr(\ln Y_N\leq-t)
=\Pr(M\leq r)
\leq\frac12 e^{-2t}+\frac12\frac{e^{-t}}{\sqrt N}
\leq e^{-t},
\]

\noindent and the inequality is trivial when $r=0$. 

\noindent For the upper tail, for $t\geq\ln2$, let $s=\lceil\sqrt N e^t\rceil$. If $s\leq N$,
the bound on $\Pr(M\geq m)$ in Eq.~\eqref{eq:end-M-tails} gives

\[
\Pr(\ln Y_N\geq t)
=\Pr(M\geq s)
\leq \exp\!\left(-\frac{e^{2t}}4\right).
\]

\noindent If $s>N$, the probability is zero. For $0\leq t<\ln2$, we use the
trivial bound $\Pr(\ln Y_N\geq t)\leq1$. 
Combining these estimates with the tail-integral identity yields

\begin{align*}
&\E[|\ln Y_N|^2]\\
&\leq2\int_0^\infty t e^{-t}\,dt
+2\int_0^{\ln2}t\,dt
+2\int_{\ln2}^\infty
t\exp\!\left(-\frac{e^{2t}}4\right)dt.
\end{align*}

\noindent The integrals are finite: the first two are elementary, while the third is bounded by an exponentially decaying
integrand, since $e^{2t}/4\geq t$ for $t\geq\ln 2$. Thus, $\E[|\ln Y_N|^2] = O(1)$, implying
$\E[|\ln Y_N|]=O(1)$ by Cauchy--Schwarz.

\noindent Since $\ln M=\tfrac12\ln N+\ln Y_N$,

\begin{equation}
\E[(\ln M)^2]
=\frac14(\ln N)^2+O(\ln N).
\label{eq:log-core-second-moment}
\end{equation}

\noindent To estimate $\mathbb{E}(\ln M\ln\ln M)$, let $f(x)=x\ln x$. For $M\geq3$, both $\ln M$ and $\tfrac12\ln N$ lie in
$[\ln3,\ln N]$ for all sufficiently large $N$. On this interval,
$|f'(x)|=|1+\ln x|=O(\ln\ln N)$. With the mean-value theorem and
$\ln M-\tfrac12\ln N=\ln Y_N$,

\[
\left|f(\ln M)-f(\tfrac12\ln N)\right|
=O(\ln\ln N)\,|\ln Y_N|.
\]

\noindent Taking expectations and using $\E[|\ln Y_N|]=O(1)$, while
$\Prb(M<3)\leq 3/N$, yields

\begin{equation*}
\begin{aligned}
\E[\ln M\ln\ln M;\,M\geq3]
&=\frac12\ln N\ln\!\left(\frac12\ln N\right)
+O(\ln\ln N)\\
&=\frac12\ln N\ln\ln N+O(\ln N).
\end{aligned}
\end{equation*}

\noindent The cases $M=1,2$ contribute only $O(1)$ to the final expectation.

\noindent Finally, averaging Eq.~\eqref{eq:fixed-core-asymptotic} over $M$ yields Eq.~\eqref{eq:critical-asymptotic}:

\begin{align*}
\E[H_N]
={}&\E[M]\ln2-\frac12\E[(\ln M)^2]\\
&+\E[\ln M\ln\ln M;\,M\geq3]+O(\ln N)\\
={}&\sqrt{\frac{\pi N}{2}}\,\ln2
-\frac18(\ln N)^2
+\frac12\ln N\ln\ln N
+O(\ln N).
\end{align*}

\subsection*{Scaling of basin entropy under loopwise additivity}
Krawitz and Shmulevich estimated the mean basin entropy by summing the
contributions of individual loops, retaining only the longest-period
attractors in each loop~\cite{krawitz2007}. Their approximation gives

\[
\mathbb E_\varepsilon[H_i\mid m_i=j]
\approx j\ln2-\ln(\sqrt{2}\,j).
\]

\noindent We now derive the corresponding asymptotic scaling of this loopwise-additive estimate,
including all states on each loop.

\noindent For a loop of length $j$, the fraction of states whose period is not
maximal is $O(2^{-j/2})$ for a positive loop and $O(2^{-j/3})$ for a
negative loop. Each such state contributes at most $\ln(2j)$ to the
difference between the logarithms of the maximal and actual periods.
Thus, including the shorter-period states changes the mean loop entropy
by $O(\ln j\,2^{-j/3})$. Averaging over the two loop signs gives

\begin{equation}
\mathbb E_{\varepsilon_i,X_i}\!\left[\ln T(X_i)\mid m_i=j\right]
=
\ln j+O(1),
\label{eq:single-loop-mean-period}
\end{equation}

\noindent where the remainder is bounded independently of $j$. Therefore, the
mean entropy of a loop of length $j$ is $j\ln2-\ln j+O(1)$.

\noindent For a uniformly random permutation in $S_m$, the expected number of
loops of length $j$ is $1/j$~\cite{arratia1992}. Hence

\[
\mathbb E_{\sigma,\varepsilon}\!\left[\sum_{i=1}^k H_i\right]
=
\sum_{j=1}^m \frac1j
\left(j\ln2-\ln j+O(1)\right).
\]

\[
\sum_{j=1}^m\frac{\ln j}{j}
=
\int_1^m\frac{\ln x}{x}\,dx+O(1)
=
\frac12(\ln m)^2+O(1)
\]

\noindent then yields

\begin{equation}
\mathbb E_{\sigma,\varepsilon}\!\left[\sum_{i=1}^k H_i\right]
=
m\ln2-\frac12(\ln m)^2+O(\ln m).
\label{eq:additive-mean-end}
\end{equation}

\bibliographystyle{unsrtnat}
\bibliography{references}

@article{kauffman1969,
  title={Metabolic stability and epigenesis in randomly constructed genetic nets},
  author={Kauffman, Stuart A},
  journal={Journal of theoretical biology},
  volume={22},
  number={3},
  pages={437--467},
  year={1969},
  publisher={Elsevier}
}

@article{kadelka2024,
  title={A meta-analysis of {B}oolean network models reveals design principles of gene regulatory networks},
  author={Kadelka, Claus and Butrie, Taras-Michael and Hilton, Evan and Kinseth, Jack and Schmidt, Addison and Serdarevic, Haris},
  journal={Science Advances},
  volume={10},
  number={2},
  pages={eadj0822},
  year={2024},
  publisher={American Association for the Advancement of Science}
}

@article{wang2012,
  title={Boolean modeling in systems biology: an overview of methodology and applications},
  author={Wang, Rui-Sheng and Saadatpour, Assieh and Albert, Reka},
  journal={Physical biology},
  volume={9},
  number={5},
  pages={055001},
  year={2012},
  publisher={IOP Publishing}
}

@article{davidich2008,
  title={Boolean network model predicts cell cycle sequence of fission yeast},
  author={Davidich, Maria I and Bornholdt, Stefan},
  journal={PloS one},
  volume={3},
  number={2},
  pages={e1672},
  year={2008},
  publisher={Public Library of Science San Francisco, USA}
}

@article{derrida1986,
  title={Random networks of automata: a simple annealed approximation},
  author={Derrida, Bernard and Pomeau, Yves},
  journal={EPL (Europhysics Letters)},
  volume={1},
  number={2},
  pages={45--49},
  year={1986}
}

@article{aldana2007,
  title={Robustness and evolvability in genetic regulatory networks},
  author={Aldana, Maximino and Balleza, Enrique and Kauffman, Stuart and Resendiz, Osbaldo},
  journal={Journal of theoretical biology},
  volume={245},
  number={3},
  pages={433--448},
  year={2007},
  publisher={Elsevier}
}

@article{balleza2008,
  title={Critical dynamics in genetic regulatory networks: examples from four kingdoms},
  author={Balleza, Enrique and Alvarez-Buylla, Elena R and Chaos, Alvaro and Kauffman, Stuart and Shmulevich, Ilya and Aldana, Maximino},
  journal={PLoS One},
  volume={3},
  number={6},
  pages={e2456},
  year={2008},
  publisher={Public Library of Science San Francisco, USA}
}

@article{daza2016basin,
  title={Basin entropy: a new tool to analyze uncertainty in dynamical systems},
  author={Daza, Alvar and Wagemakers, Alexandre and Georgeot, Bertrand and Gu{\'e}ry-Odelin, David and Sanju{\'a}n, Miguel AF},
  journal={Scientific reports},
  volume={6},
  number={1},
  pages={31416},
  year={2016},
  publisher={Nature Publishing Group UK London}
}

@article{daniels2018,
  title={Criticality distinguishes the ensemble of biological regulatory networks},
  author={Daniels, Bryan C and Kim, Hyunju and Moore, Douglas and Zhou, Siyu and Smith, Harrison B and Karas, Bradley and Kauffman, Stuart A and Walker, Sara I},
  journal={Physical review letters},
  volume={121},
  number={13},
  pages={138102},
  year={2018},
  publisher={APS}
}

@article{kadelka2026canalization,
  title={Canalization as a stabilizing principle of gene regulatory networks: a discrete dynamical systems perspective},
  author={Kadelka, Claus},
  journal={NPJ Systems Biology and Applications},
  volume={12},
  number={1},
  pages={30},
  year={2026},
  publisher={Nature Publishing Group UK London}
}

@article{krawitz2007,
  title = {Basin Entropy in {B}oolean Network Ensembles},
  author = {Krawitz, Peter and Shmulevich, Ilya},
  journal = {Phys. Rev. Lett.},
  volume = {98},
  issue = {15},
  pages = {158701},
  numpages = {4},
  year = {2007},
  month = {Apr},
  publisher = {American Physical Society},
  doi = {10.1103/PhysRevLett.98.158701},
  url = {https://link.aps.org/doi/10.1103/PhysRevLett.98.158701}
}

@article{flyvbjerg1988,
  title={Exact solution of {K}auffman's model with connectivity one},
  author={Flyvbjerg, Henrik and Kjaer, NJ},
  journal={Journal of Physics A: Mathematical and General},
  volume={21},
  number={7},
  pages={1695--1718},
  year={1988}
}

@article{drossel2005,
  title={Number and length of attractors in a critical {K}auffman model with connectivity one},
  author={Drossel, Barbara and Mihaljev, Tamara and Greil, Florian},
  journal={Physical review letters},
  volume={94},
  number={8},
  pages={088701},
  year={2005},
  publisher={APS}
}

@article{samuelsson2005,
  title={Random maps and attractors in random {B}oolean networks},
  author={Samuelsson, Bj{\"o}rn and Troein, Carl},
  journal={Physical Review E—Statistical, Nonlinear, and Soft Matter Physics},
  volume={72},
  number={4},
  pages={046112},
  year={2005},
  publisher={APS}
}

@article{fink2023,
  title={Number of attractors in the critical {K}auffman model is exponential},
  author={Fink, TMA and Sheldon, FC},
  journal={Physical Review Letters},
  volume={131},
  number={26},
  pages={267402},
  year={2023},
  publisher={APS}
}

@article{fink2024,
  title={Exact behavior of the critical {K}auffman model with connectivity one},
  author={Fink, TMA},
  journal={Physical Review Research},
  volume={6},
  number={4},
  pages={043315},
  year={2024},
  publisher={APS}
}

@article{zacharovas2004,
  title={Distribution of the logarithm of the order of a random permutation},
  author={Zacharovas, V},
  journal={Lithuanian Mathematical Journal},
  volume={44},
  number={3},
  pages={296--327},
  year={2004},
  publisher={Springer}
}

@article{mihaljev2006,
  title = {Scaling in a general class of critical random {B}oolean networks},
  author = {Mihaljev, Tamara and Drossel, Barbara},
  journal = {Phys. Rev. E},
  volume = {74},
  issue = {4},
  pages = {046101},
  numpages = {10},
  year = {2006},
  month = {Oct},
  publisher = {American Physical Society},
  doi = {10.1103/PhysRevE.74.046101},
  url = {https://link.aps.org/doi/10.1103/PhysRevE.74.046101}
}

@misc{bavisetty2026,
      title={Permutation theory governs long-term dynamics of critical {B}oolean networks}, 
      author={Venkata Sai Narayana Bavisetty and Matthew Wheeler and Julian Vignes and Jones, Jr., Matthew Stephen and Ruodan Liu and Sean Campbell and Claus Kadelka},
      year={2026},
      eprint={2608.07790},
      archivePrefix={arXiv},
      primaryClass={q-bio.MN},
      url={https://arxiv.org/abs/2608.07790}, 
}

@article{arratia1992,
 ISSN = {00911798, 2168894X},
 URL = {http://www.jstor.org/stable/2244659},
 author = {Richard Arratia and Simon Tavaré},
 journal = {The Annals of Probability},
 number = {3},
 pages = {1567--1591},
 publisher = {Institute of Mathematical Statistics},
 title = {The Cycle Structure of Random Permutations},
 urldate = {2026-10-16},
 volume = {20},
 year = {1992}
}

@article{flajolet1995,
title = {On {Ramanujan}'s {Q}-function},
journal = {Journal of Computational and Applied Mathematics},
volume = {58},
number = {1},
pages = {103-116},
year = {1995},
issn = {0377-0427},
doi = {https://doi.org/10.1016/0377-0427(93)E0258-N},
url = {https://www.sciencedirect.com/science/article/pii/0377042793E0258N},
author = {Philippe Flajolet and Peter J. Grabner and Peter Kirschenhofer and Helmut Prodinger}
}

\end{document}